\documentclass[conference]{IEEEtran}
\IEEEoverridecommandlockouts
\usepackage{amsmath,amssymb,amsfonts}
\usepackage{algorithmic}
\usepackage{graphicx}
\usepackage{textcomp}
\usepackage{cleveref}
\usepackage{xcolor}
\usepackage{placeins}
\usepackage{stfloats}  
\usepackage[style=numeric,sorting=none,maxnames=6, minnames=1]{biblatex}
\begin{document}

\title{PyPSA-DE: Open-source German energy system model reveals savings from integrated planning}

\author{\IEEEauthorblockN{Michael Lindner, Julian Geis, Toni Seibold, Tom Brown }
\IEEEauthorblockA{\textit{Department of Digital Transformation in Energy Systems} \\
\textit{Technical University of Berlin}\\
10587 Berlin, Germany \\
corresponding author: m.lindner@tu-berlin.de}
}

\IEEEpubid{\makebox[\textwidth][l]{%
\parbox{\textwidth}{
\vspace{9\baselineskip}
\footnotesize
\centering
979-8-3315-1278-1/25/\$31.00~\copyright~2025 IEEE. Personal use of this material is permitted. Permission from IEEE must be obtained for all other uses, in any current or future media, including reprinting/republishing this material for advertising or promotional purposes, creating new collective works, for resale or redistribution to servers or lists, or reuse of any copyrighted component of this work in other works. DOI: 10.1109/EEM64765.2025.11050093
}}}
%\hspace{\columnsep}\makebox[\columnwidth]{}
% \IEEEpubid{
% \footnotesize
% 979-8-3315-1278-1/25/\$31.00~\copyright~2025 IEEE. Personal use of this material is permitted. Permission from IEEE must be obtained for all other uses, in any current or future media, including reprinting/republishing this material for advertising or promotional purposes, creating new collective works, for resale or redistribution to servers or lists, or reuse of any copyrighted component of this work in other works.
% }
% \and
% \and
% \and

%\thanks{979-8-3315-1278-1/25/\$31.00~\copyright~2025 IEEE. Personal use of this material is permitted. Permission from IEEE must be obtained for all other uses, in any current or future media, including reprinting/republishing this material for advertising or promotional purposes, creating new collective works, for resale or redistribution to servers or lists, or reuse of any copyrighted component of this work in other works.} 

\maketitle

%\IEEEpubid{\makebox[\columnwidth]{979-8-3315-1278-1/25/\$31.00~\copyright~2025 IEEE \hfill} \hspace{\columnsep}\makebox[\columnwidth]{}}

\begin{abstract}
    % Germany has set an ambitious decarbonization target of reaching net zero emissions by 2045. In this contribution we explore how integrated cross-sectoral system planning can lower costs compared to existing national plans. Our new framework PyPSA-DE simulates supply, demand, storage and transmission networks to cover energy and non-energy demand in the electricity, heating, transport, agriculture, waste and industry sectors. It builds a linear optimization problem to minimize total system costs of the energy system infrastructure in Germany and its neighboring countries, with high spatial and temporal resolution. 
    
    % While the model shows strong transmission grid development, primarily to connect wind in the North to demand in the South, total expansion is up to one third lower than in the national grid development plan, lowering costs by 80 bn € to  203 bn €. These savings contribute to an overall reduction in electricity prices, and are mainly due to fully integrated planning and operation of the hydrogen and electricity systems, a market design with regional price zones, and a system-optimal usage of offshore wind resources. PyPSA-DE is fully open-source and can readily be adapted to study further pressing issues around the energy transition. 

    %Shorter

    Germany has set an ambitious target of reaching net zero greenhouse gas emissions by 2045. We explore how integrated cross-sectoral planning can reduce costs compared to existing national plans. Our new linear optimization model PyPSA-DE simulates the electricity and hydrogen transmission networks, as well as supply, demand, and storage in all sectors of the energy system in Germany and its neighboring countries with high spatial and temporal resolution.
    While our new model shows strong electricity transmission grid development, total expansion is one third lower than in the national grid development plan, lowering costs by 92 billion €\textsubscript{2020} to 191 billion €\textsubscript{2020} and average grid tariffs by 7.5 €\textsubscript{2020}~/~MWh.
    %so that prices reduce in all regions of Germany, but most strongly in the North
    These savings are mainly due to integrated planning and operation, a market design with regional prices, and a system-optimal usage of offshore wind.
    %Savings for the hydrogen network are more modest.
    PyPSA-DE is open-source and can readily be adapted to study related issues around the energy transition.  
    
\end{abstract}


\section{Introduction}
Germany has set an ambitious target of reaching net zero emissions by 2045. This long-term commitment is underpinned by laws addressing specific sectors of the energy system, such as build-out targets for renewable energy sources (RES), the phase-out of coal power, a ban on fossil boilers in individual heating, investments into a new hydrogen transmission grid and ambitious plans for power grid expansion, to name just a few. While the implemented measures are projected by the German federal environmental agency (UBA) to significantly reduce emissions,  they are not deemed sufficient to reach net-zero \cite{UBA2024}.  
Against this backdrop we have developed PyPSA-DE  - the first open-source cross-sectoral high-resolution model of the German energy system. It allows for an integrated planning of the whole energy system, uncovering synergies arising from sector-coupling that may help to achieve climate neutrality in a cost-optimal way. While the model already represents a wide-range of policies and technology options, the fully open-source publication of its input data and modeling workflow ensures that it can be readly adapted by diverse actors to study the impact of new political interventions or technological innovations. In this conference paper, the default modeling assumptions of PyPSA-DE are outlined and selected modeling results and their implications for the German energy transition are discussed.

%\IEEEpubidadjcol

\section{Methods}

We developed PyPSA-DE on the basis of the widely used European energy system model PyPSA-Eur~\cite{PyPSAEur,PyPSAEurSec,SpeedTechnological2022} to resolve the German system in high spatial, temporal and sectoral detail. PyPSA-DE simulates supply, demand, storage and transmission networks to cover energy and non-energy demand in the electricity, heating, transport, agriculture, waste and industry sectors. In 5-year steps with myopic foresight from 2020 to 2045, PyPSA-DE builds a linear optimization problem to minimize total system costs of the energy system infrastructure in Germany and its neighboring countries, with variable spatial and temporal resolution across full weather years. Among other things, PyPSA-DE computes a cost-optimal allocation of renewables in regional scope, cross-border trade of electricity and renewable fuels, expansion of hydrogen and electricity networks, and adaptation of different heating and thermal energy storage technologies in district as well as individual heating. The grid is modeled with an optimal linearized power flow with piecewise linear losses, an approach found to agree well with the full load flow calculation~\cite{neumann2022assessments}. The base networks is extracted from openstreetmaps and validated against official data~\cite{xiong2025modelling}.

The new model PyPSA-DE implements Germany-specific policies such as the nuclear and coal phase-out laws, or the net-zero target for 2045, as well as EU-wide regulations and CO$_2$ emissions budgets. PyPSA-DE augments the PyPSA-EUR dataset with several Germany-specific data sources, including the national grid development plans for electricity~\cite{bundesnetzagentur_bedarfsermittlung_2024} and hydrogen~\cite{bundesnetzagentur_genehmigung_2024}, and the register of power and gas units in Germany~\cite{MaStR}.
Furthermore, the model has been validated against historical data of the German electricity system for 2020. 
While PyPSA-DE can be configured in many different ways, the results discussed in this article were obtained for a net-zero scenario, with 3-hourly resolution and 49 regions (30 regions for Germany, and 19 regions for the European neighbors). 

\section{Results}

\subsection{Expansion of renewable energy sources and CO$_2$ emissions.} 
\begin{figure}
    \centering
    \includegraphics[width=\linewidth]{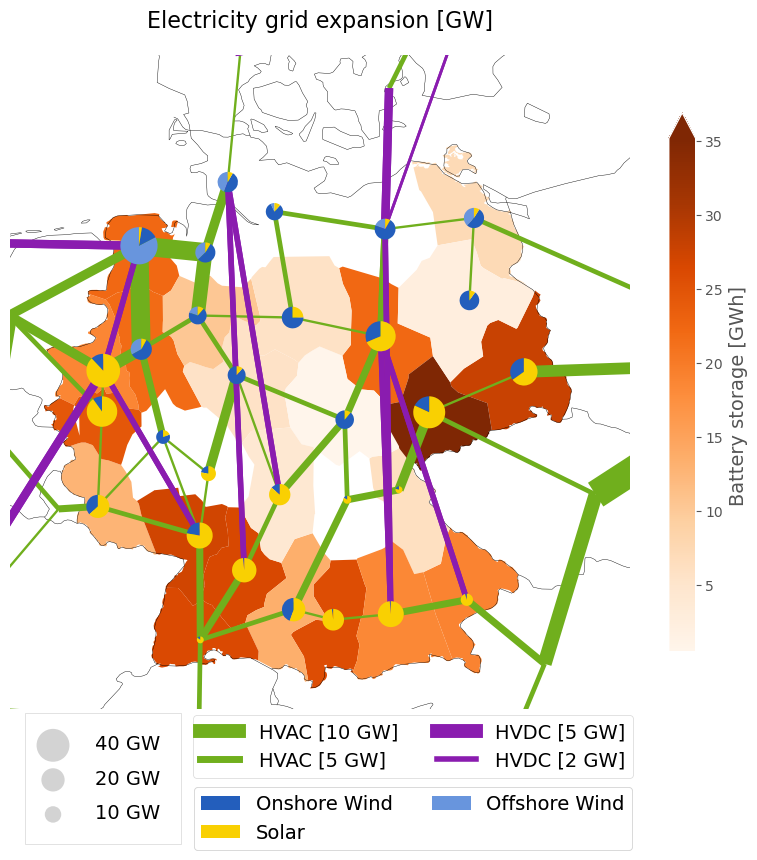}
    \caption{Installed capacities of renewable energy sources (RES) and batteries, and transmission grid expansion for the year 2045 using historical weather data from 2019. Offshore wind is shown at its grid connection point. The installed capacities follow the renewable potential, with onshore wind found predominantly in the North and solar power found predominantly in the South.}
    \label{fig:capacities}
\end{figure}

The model results indicate that the electricity sector, will become the backbone of the future energy system in Germany, and that the electricity supply will be dominated by renewable energy sources (RES). These are predominantly expanded where the expected yields are highest, e.g., for onshore wind in the North of Germany, see~\Cref{fig:capacities}. Only after land use constraints are hit for that technology, are less profitable sites in the South utilized. Average wholesale electricity prices are expected to slightly decrease from their current high point and to stabilize from 2030 onwards at around 80€/MWh. The total investment needs for the German electricity system in the period 2025-2045 exceed 730 billion €\textsubscript{2020}, of which the largest share (around 60\%, or 425 billion €\textsubscript{2020}) goes to the expansion of wind and solar, followed by investments into the transmission grid, which total 191 billion €\textsubscript{2020} in PyPSA-DE. A detailed representation of the electricity distribution grid is outside the scope of the model, but recent studies estimate investment needs of a magnitude similar to that of the transmission grid~\cite{fraunhofer_isi_langfristszenarien_2024,energie__management_verteilnetzausbau_2024}, so that the real total will be even greater than 730 billion euros\textsubscript{2020}.

\subsection{Backup generation and flexibility.}
\begin{figure*}[b!]
    \centering
    \includegraphics[width=0.8\textwidth]{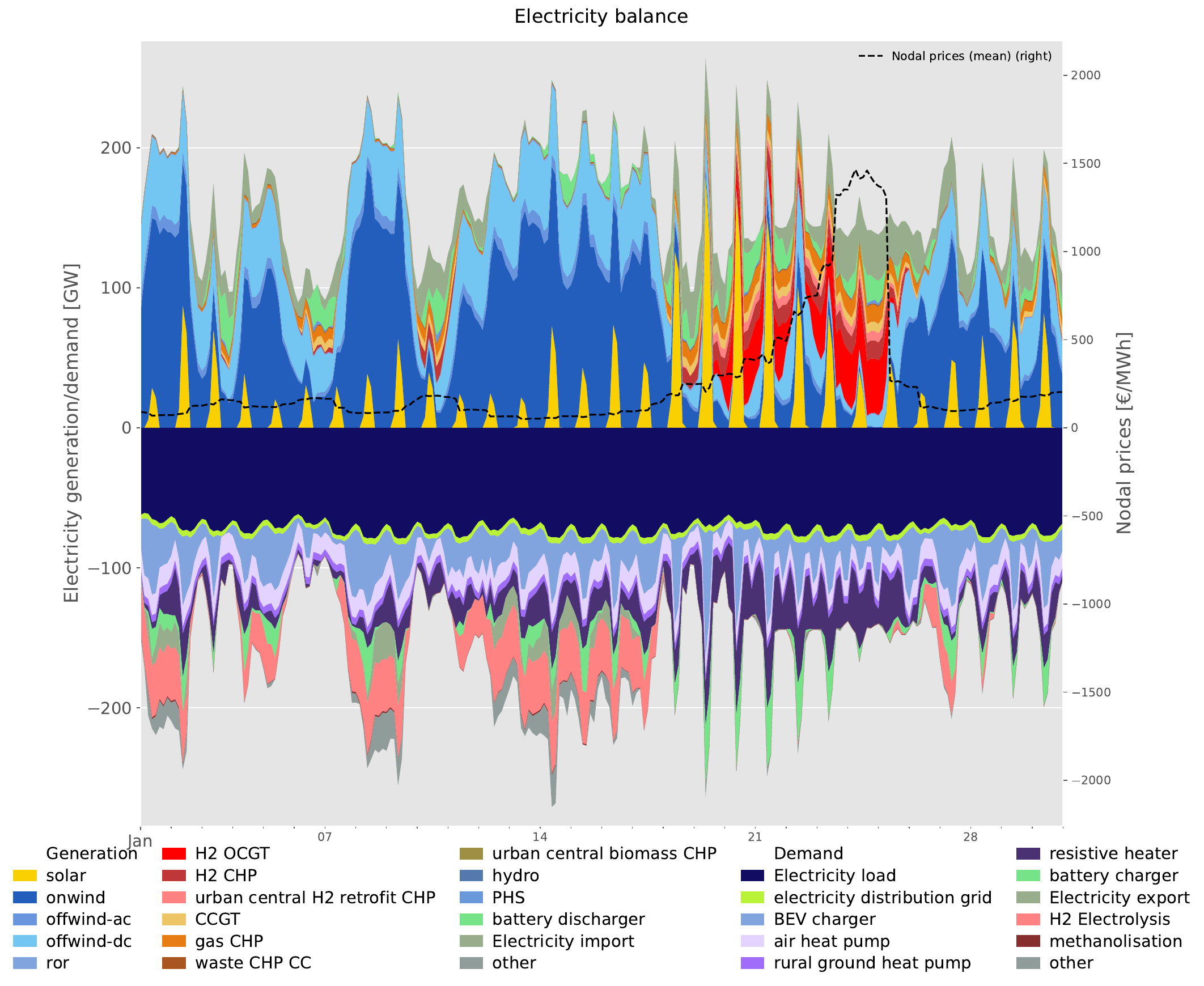}
    \caption{Electricity balance for the month of January for the year 2045. During the event of low renewable energy generation (19th to 26th), hydrogen and gas backup power plants step in, supplemented by electricity imports.}
    \label{fig:balance}
\end{figure*}

\Cref{fig:balance} shows an example electricity balance of the decarbonized energy system in 2045 for the month of January, which contains a period with low RES generation. During this week-long event the backup power plants (68 GW of hydrogen and 19 GW of older unabated natural gas plants) are running at capacity, supplemented by power imports from the neighboring countries. In general, the integration into the European electricity system proves to be vital for balancing, with gross electricity exchange exceeding by far the net imports. During the low-RES event the production of hydrogen, which usually absorbs the excess power in wind-rich periods, is halted. Aside from providing temporal flexibility to the system, the electrolysis in Germany has another vital role, namely integrating offshore wind energy, without the need for expensive underground DC cables\footnote{Cheaper, overhead DC cables have faced considerable public opposition, such that, the German parliament mandated that all future DC project shall be underground.}. As a consequence the electrolysis is predominantly situated near the coast.

Using the methodology of Artelys~\cite{artelys_artelys_2023} we can assess the flexibility needs at different time scales (daily and weekly). The methodology works by quantifying deviations of the residual load from means taken over periods of different lengths, then adjusting the residual load by the contribution of each flexibility technology. Taking the daily flexibility as an example, the positive deviation of the residual load from its daily average is quantified, then as the contribution of batteries or demand-side management is added to the residual load, the deviation is measured again. Eventually once all technologies are considered, the residual load drops to zero, since the system is balanced, and the exercise is complete.

\Cref{fig:flex} shows which flexibility resources provide this flexibility. At the daily scale the needs rise by a factor 10 betweeen 2020 and 2045, primarily to integrate solar fluctuations, and are met by a mix of electric vehicle charging management and stationary batteries. At the weekly scale the needs grow by a factor 5, supplied mainly by trade and electrolysis.

\subsection{Integrated planning with regional prices saves on transmission costs.} 
\begin{figure*}
    \centering
    \includegraphics[width=0.4\textwidth]{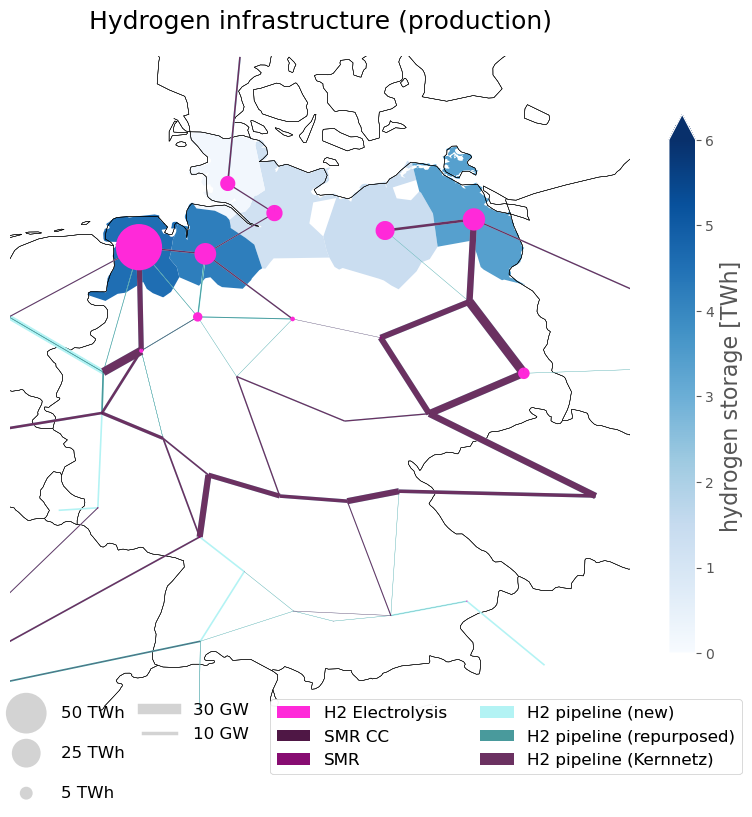}
    \includegraphics[width=0.4\textwidth]{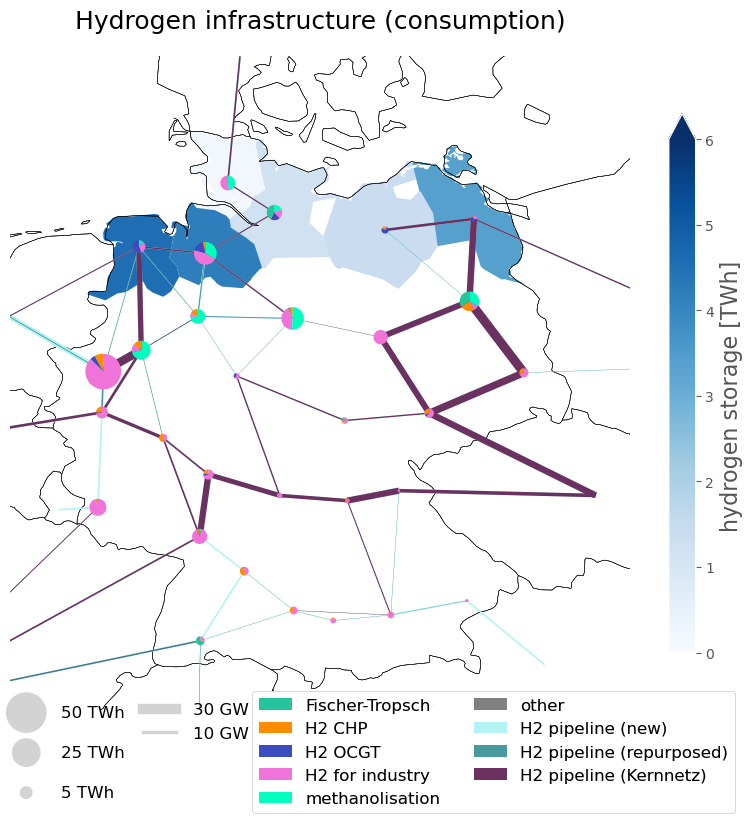}

    \caption{Expansion of the hydrogen grid until 2045, shown together with locations of H2 production (left) and H2 demand (right). The electrolysis is placed primarily near the coastline to help with the integration of offshore wind resources and thus save on expensive DC projects.}
    \label{fig:h2}
\end{figure*}

While the model shows strong transmission grid development, especially in the period 2025-2035, it identifies the potential to reduce grid development by up to one third compared to the national grid development plan (NDP)~\cite{bundesnetzagentur_bedarfsermittlung_2024}, lowering costs by 92 billion €\textsubscript{2020} to  191 billion €\textsubscript{2020}. These savings are mainly due to a fully integrated planning and operation, which promotes the system-friendly location and dispatch of flexibility technologies like electrolysis and utility-scale batteries, as well as due to cost-optimal usage of offshore wind resources and regional price zones.  While the NDP expands the grid to remove congestion arising from the market dispatch of a single bidding zone, leading to substantial grid expansion, we use regional prices to manage congestion within Germany. Regional prices help the system to dispatch local assets to manage congestion, leading to less expansion as well as congestion rents that lower the grid tariffs charged to consumers. Furthermore regional prices allow the market to guide the electrolysis capacities to the coast, where they can absorb offshore wind power (see \Cref{fig:h2}) - a result that was also found by \cite{vom_scheidt_integrating_2022}. 

\begin{figure*}
    \centering
    \includegraphics[width=0.8\textwidth]{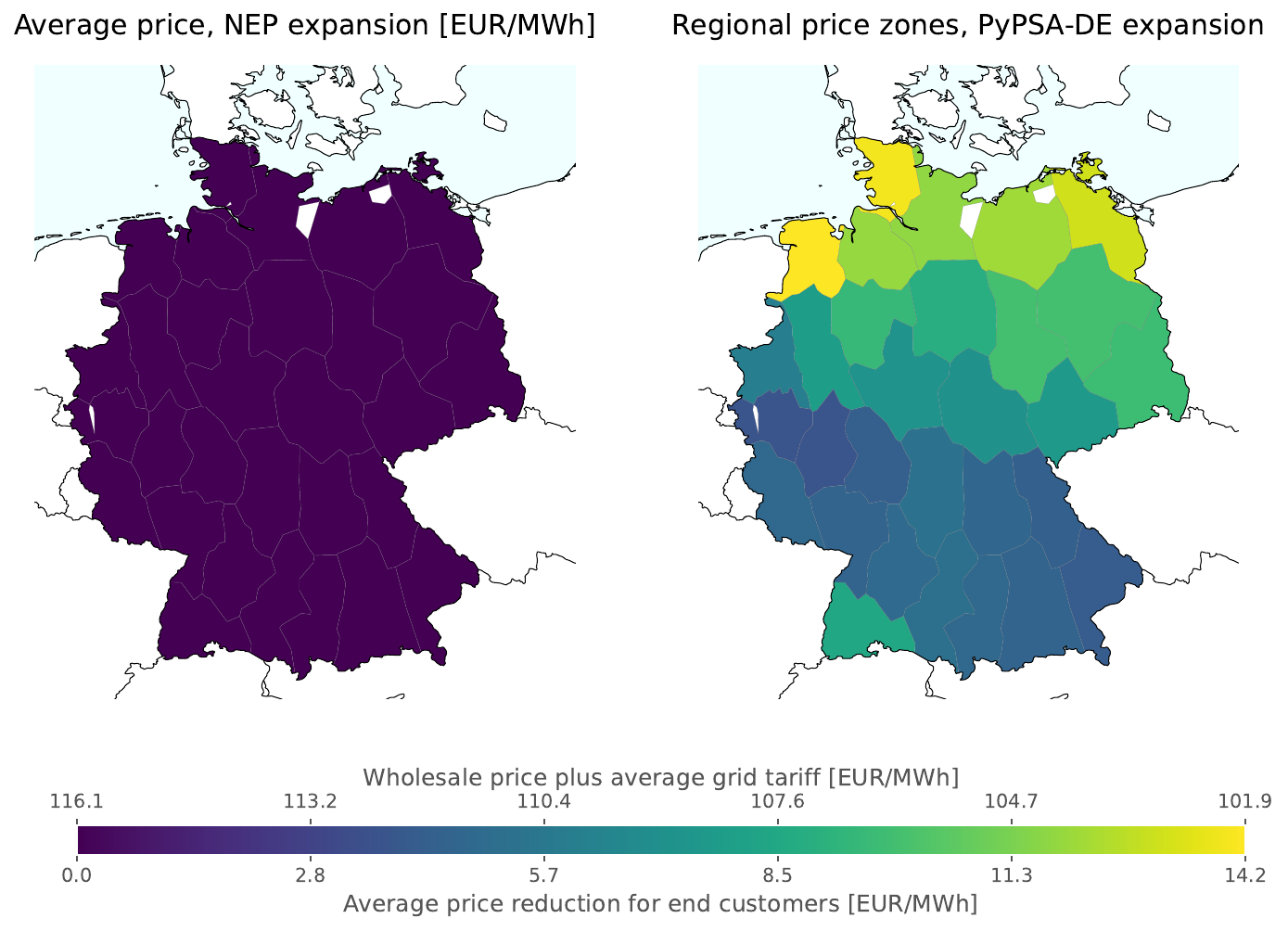}
    \caption{Comparison of the average wholesale electricity price plus average grid tariffs in 2045. Left: scenario with a single bidding zone and grid expansion according to the national grid development plan (NEP), Right: scenario with regional price zones and reduced grid development according to PyPSA-DE. In the PyPSA scenario all end users benefit from a lower average electricity price, despite price differences between the regional zones.}
    \label{fig:prices}
\end{figure*}

The lower grid development reduces grid tariffs by 7.5 €/MWh on average compared to the NDP plan. Due to the regional prices, there are variations in the wholesale price from region to region, but because of the lower grid tariffs, the end consumer prices reduce in all regions, varying from 3.5 €/MWh in the south down to 14.2 €/MWh in the north near the coast (see \Cref{fig:prices}). A recent study on locational prices in GB led to a similar conclusion, finding lower consumer prices in all regions~\cite{fti_consulting_assessment_2023}.

As a linear model, PyPSA-DE has to make certain simplifying assumptions, e.g., the complete, non-linear load flow is approximated with a linear load flow, with piecewise linear losses~\cite{neumann2022assessments}. The linear load flow is solved with a computationally efficient cycle-based formulation~\cite{horsch2018linear}, implemented in the python package PyPSA~\cite{PyPSA}. Furthermore, based on~\cite{shokri_gazafroudi_topology-based_2022}, the $N-1$ criterium is approximated by restricting the maximal line loading to 70\%. New lines are not build as discrete units, but optimized continuously and discretized in a further step as soon as a certain maximum loading is exceeded~\cite{neumann2022assessments}. Because of such simplifications, our model is not a substitute for the full grid planning of the network operators, but it is indicative how an integrated planning approach with regional price signals can reduce costs.

\subsection{Hydrogen infrastructure.} Next to the expansion of the electricity grid, the model constructs a hydrogen pipeline network that connects electrolysis, hydrogen imports from neighboring countries and large-scale consumers in the industry sector, as well as in e-fuel production and backup power generation. Following the approach in~\cite{PyPSAEurSec} pipelines are modeled as linear transport links, without explicit consideration of gas flow physics. We find that the official network development plans~\cite{bundesnetzagentur_genehmigung_2024} are sufficient for the transport demand in 2045. Even a hydrogen network with less transport capacity might be enough to serve a comparatively low total hydrogen demand of 191 TWh, unlocking potential cost savings. In particular, with 130 TWh, the largest share of hydrogen is produced domestically, highlighting the importance of electrolysis for the electricity system, which serves as a source of flexibility (see~\Cref{fig:flex}) and helps integrate offshore wind resources (see~\Cref{fig:h2}).

\section{Conclusion} 

Until now much of the energy system planning has been done in separate silos for electricity, natural gas and hydrogen. By pursuing an integrated cross-sectoral planning approach, as well as considering changes to the market structure like regional pricing, we were able to identify significant cost savings. Regional prices allow the system to manage congestion with flexibility, electrolysers can be placed at the coast to absorb offshore wind, and many of the planned long HVDC cables from North to South can be avoided.

By publishing our model online under an open-source license~\cite{pypsa_de} we hope to inspire its further use by other researchers. It is already being adapted to study further pressing topics of the Germany energy transition, e.g., the impact of decarbonization on the industry sector, price-formation in an all-renewable energy system, and the benefits of thermal energy storage. Furthermore, it may serve as a prototype for building similar country specific open-source models within Europe on the basis of PyPSA-Eur. In the context of the Ariadne project, PyPSA-DE  is provides an important independent contribution to research on German energy policy~\cite{luderer2025energiewende}.

% \bibliographystyle{plainnat}
% \bibliography{references} 
\printbibliography

\appendix

\section*{Additional figures}

\begin{center}
    (see next pages)
\end{center}

\begin{figure*}
    \centering
    \includegraphics[width=0.7\textwidth]{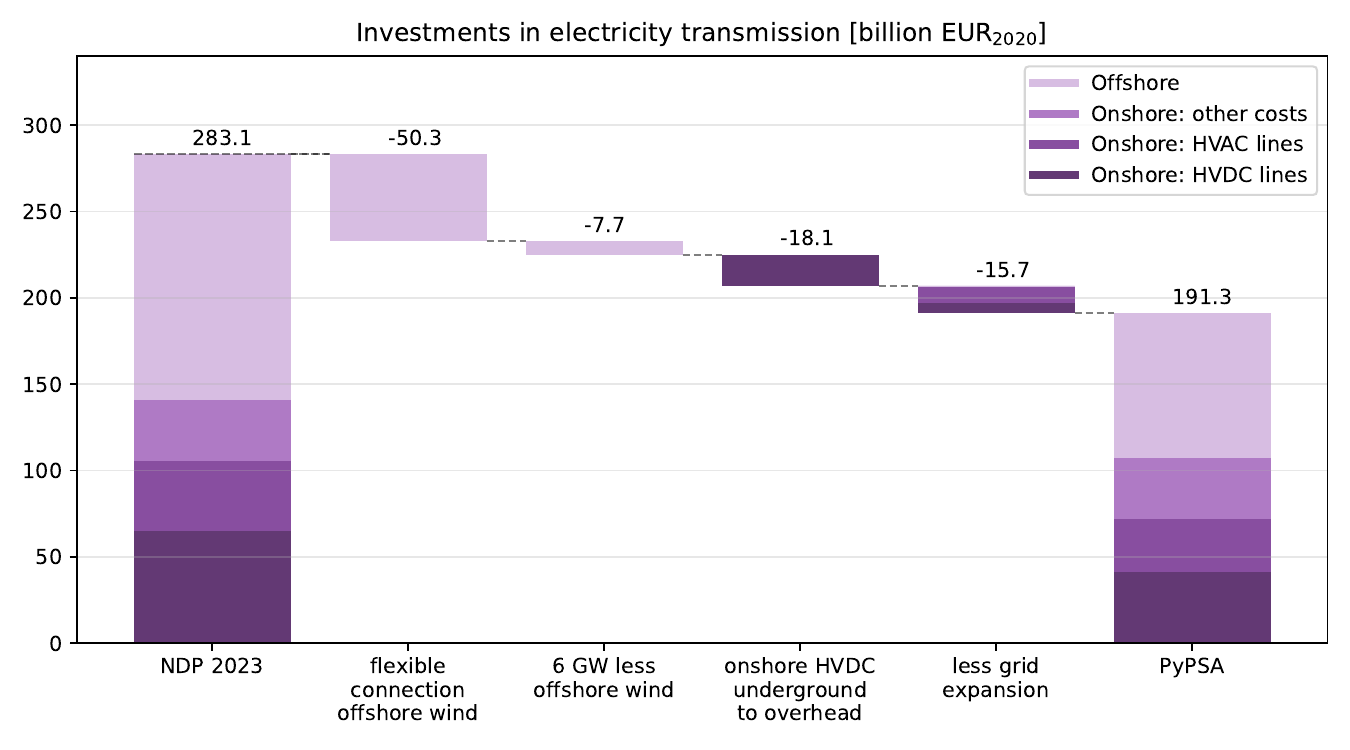}
    \caption{Investment needs for the electricitry transmission grid until 2045 according to the NEP, and potential savings according to the PyPSA-DE methodology, differentiated by enabling assumption and affected grid components.}
    \label{fig:investment}
\end{figure*}

\begin{figure*}[!b]
    \centering
    \includegraphics[width=0.7\linewidth]{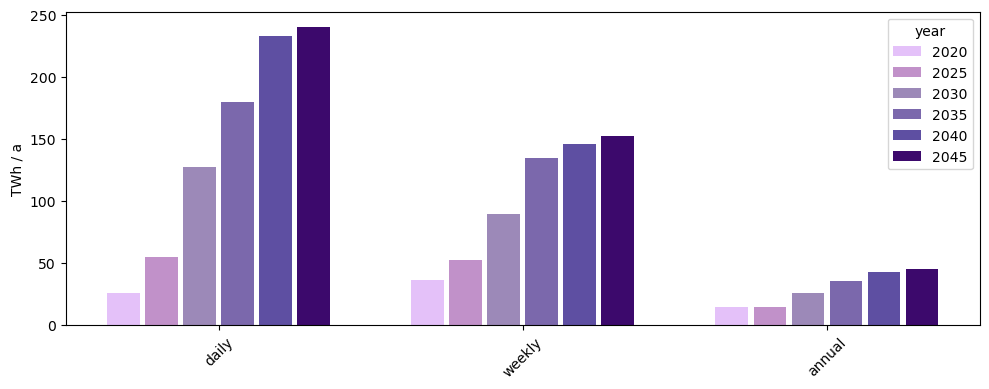}
    \includegraphics[width=0.7\linewidth]{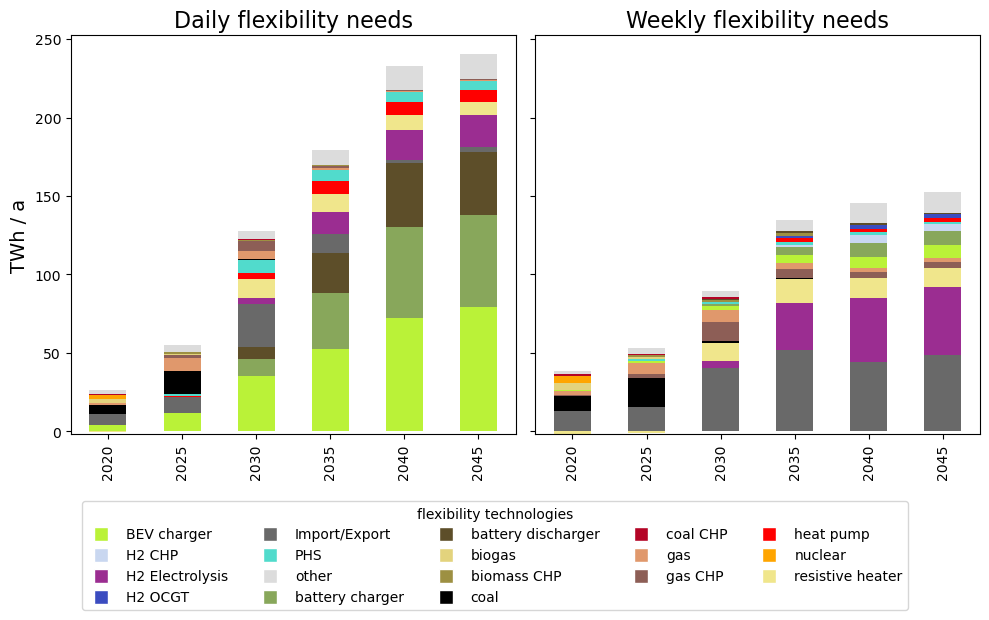}

    \caption{Top: Total daily, weekly and yearly flexibility needs for each model year. Bottom: Technologies that cover the daily and weekly flexibility needs. To compute the flexibility needs the method of Artelys is used
    ~\cite{artelys_artelys_2023}.}
    \label{fig:flex}

\end{figure*}

\end{document}